\pdfoutput=1

\documentclass[referee]{raa}            % referee version: for submission

\usepackage{graphicx,times}             %for PS/EPS graphics inclusion, new
\usepackage{natbib}
\usepackage{amssymb,amsmath}
\bibpunct{(}{)}{;}{a}{}{,}
\usepackage[a4paper=true,pagebackref=true]{hyperref}
\hypersetup{colorlinks = true, linkcolor = green, anchorcolor = red, citecolor = blue, filecolor = red, pagecolor = red, urlcolor = red}

\begin{document}

   \title{The afterglow emission from a  stratified jet in GRB 170817A
%\footnotetext{$*$ Supported by the National Natural Science Foundation of China.}
}
%   \subtitle{I. Place Your Subtitle Here}

   \volnopage{Vol.0 (20xx) No.0, 000--000}      %%preserved for Editor. DOn't remove!
   \setcounter{page}{1}          %%starting page, preserved for Editor. DOn't remove!

   \author{Kang-Fa Cheng
      \inst{1,2}
   \and Xiao-Hong Zhao
      \inst{1,3,4}
  \and Bin-Bin Zhang
     \inst{5,6,7}
   \and Jin-Ming Bai
      \inst{1,3,4}
   }

   \institute{Yunnan Observatories, Chinese Academy of Sciences, 650216, Kunming, China; {\it zhaoxh@ynao.ac.cn}\\
        \and
             University of the Chinese Academy of Sciences, Yuquan Road 19, Shijingshan Block, 100049, Beijing, China\\ 
       \and  
             Center for Astronomical Mega-Science, Chinese Academy of Sciences, Beijing, China\\
       \and
             Key Laboratory for the Structure and Evolution of Celestial Objects, Chinese Academy of Sciences, Kunming, China\\
       \and
             School of Astronomy and Space Science, Nanjing University, Nanjing 210093, China\\
       \and
             Key Laboratory of Modern Astronomy and Astrophysics (Nanjing University), Ministry of Education, China\\
       \and
             Department of Physics and Astronomy, University of Nevada Las Vegas, NV 89154, USA\\
\vs\no
   {\small Received~~20xx month day; accepted~~20xx~~month day}}

\abstract{ The afterglow of GRB 170817A has been detected for more than three years, but the origin of the multi-band afterglow light curves remains under debate.
A classical top-hat jet model is faced with difficulties in producing a shallow rise of the afterglow light curves as observed $(F_{\nu} \propto T^{0.8})$. Here we reconsider the model of stratified ejecta with energy profile of $E(>\Gamma \beta)=E_0(\Gamma \beta)^{-k}$ as the origin of the afterglow light curves of the burst, where $\Gamma$ and $\beta$ are the Lorentz factor and speed of the ejecta, respectively. $k$ is the power-law slope of the energy profile. We consider the ejecta are collimated into jets. Two kinds of jet evolutions are investigated, including a lateral-spreading jet and a non-lateral-spreading jet. We fit the multi-band afterglow light curves, including the X-ray data at one thousand days post-burst, and find that both the models of the spreading and non-spreading jets can fit the light curves well, but the observed angular size of the source and the apparent velocity of the flux centroid for the spreading jet model are beyond the observation limits, while the non-spreading jet model meets the observation limits. Some of the best-fit parameters for the non-spreading jet model, such as the number density of the circumburst medium $\sim10^{-2}$ cm$^{-3}$ and the total jet kinetic energy $E \sim 4.8\times 10^{51}$ erg, also appear plausible. The best-fit slope of the jet energy profile is $k \sim 7.1$. Our results suggest that the afterglow of GRB 170817A may arise from the stratified jet and that the lateral spreading of the jet is not significant. 
}

   \authorrunning{K.-F. Cheng et al.}            %author_head in even pages
   \titlerunning{The afterglow emission from a  stratified jet in GRB 170817A}  % title_head in odd pages

   \maketitle
%% The author head (on even pages) and the title head (on odd pages) will be
%% automatically extracted from \author{} and \title{}. Whenever the title is too long,
%% you will be asked to supply a shorter one by inserting either \authorrunning{} or
%% \titlerunning{} before \maketitle. Anyway, you can specify your own heads.
%%
%%
%% Note: In the following text body of your manuscript, please note several differences from
%%       other major journals:
%% (1) \subsection{Please Capitalize the First Letter of Each Notional Word in Subsection Title}
%% (2) Please Capitalize the First Letter of Each Notional Word in all tables' captions

%
%________________________________________________ sections below
%
\section{Introduction}           %% first-level sections will be auto-capitalized
\label{sect:intro}

GRB 170817A is known as the electromagnetic counterpart of the first binary neutron stars merger event.
It was triggered by Fermi Gamma-Ray Burst Monitor $\sim 1.7$ seconds after the triggered of gravitational waves (GW 170817) (e.g., \citealt{Abbott+etal+2017a, Abbott+etal+2017b}). Later, the X-ray, radio, and optical afterglow of GRB 170817A were detected in sequence on $\sim 9$, $\sim16$, and $\sim 110$ days post-merger (e.g., \citealt{Alexander+etal+2017, Haggard+etal+2017, Hallinan+etal+2017, Lyman+etal+2018, Margutti+etal+2017, Margutti+etal+2018, Troja+etal+2017}). Actually, the related optical emission was detected earlier at $\sim 15$ days post-merger, but it was thought to be the kilonova emission (e.g., \citealt{Abbott+etal+2017b, Arcavi+etal+2017, Kasliwal+etal+2017, Pian+etal+2017}).
GRB 170817A is an unconventional short gamma-ray burst (SGRB) which is reflected in two aspects. First, the prompt emission is low-luminosity and the isotropic equivalent energy is $E_{\gamma,iso}\sim6\times10^{46}$ erg (e.g., \citealt{Margutti+etal+2018}).
Such an isotropic equivalent energy is $\sim$ four orders of magnitude lower than the classical SGRBs for which the typical energy is $E_{\gamma,iso}\sim10^{50} - 10^{52}$ erg (\citealt{Berger+2014, Fong+etal+2015}). Second, the afterglow light curves of the burst have a super-long and slowly rising phase (with $F_{\nu} \propto T^{0.8}$ ) with a duration of $\sim 160$ days (\citealt{Mooley+etal+2018a}). 

Given the peculiar properties of the burst, there are mainly four potential scenarios. First, a classical top-hat jet (THJ) is viewed on-axis. This scenario needs a low-energy jet to produce the low-luminosity gamma-ray emission in the prompt phase, but such a low-energy jet is unlikely to break out the ejecta whose mass is suggested to be $\thickapprox0.05M_{\odot}$ by the observed ultraviolet-optical-infrared counterpart (\citealt{Kasliwal+etal+2017}).  In addition, this scenario cannot produce the slowly rising phase of the afterglow. Second, a classical THJ is viewed off-axis. The slowly rising phase of afterglow has ruled out this scenario, since an off-axis observed THJ will lead to a rise with lope $\gtrsim 3$, much steeper than the observed $F_{\nu} \propto T^{0.8}$ (e.g., \citealt{Lazzati+etal+2018, Margutti+etal+2017, Mooley+etal+2018a}).

The other two scenarios involve the radial and angular structures of ejecta. Third, a quasi-spherical shell with energy injection model (hereafter 'QSSEI').
The QSSEI model has a radial profile of energy, which is distributed as $E(>u) \propto u^{-k}$ (e.g., \citealt{Gill+etal+2018, Mooley+etal+2018a, Huang+Li+2018}), where $u=\Gamma \beta$ is the proper velocity of the ejecta and $k$ is the energy injection index.
Such energy distribution indicates that there is more energy residing in the slower ejecta.
As the fastest ejecta gradually be decelerated by the external medium, the slower and more energetic ejecta will catch up with it and energize it (e.g., \citealt{Sari+Meszaros+2000, Nakamura+Shigeyama+2006, Gill+etal+2018}).
This scenario can explain the slowly rising phase of the afterglow
but it is hard to produce a sharply decaying observed flux density as $F_{\nu} \propto T^{-2.2}$ (\citealt{Mooley+etal+2018c, Mooley+etal+2018b}).
Moreover, by taking a global network of 32 radio telescopes, \cite{Ghirlanda+etal+2019} has constrained the apparent source size to be smaller than 2.5 milliarcseconds at $90\%$ confidence level and the isotropic outflow scenario has been ruled out.

The fourth scenario is a wide-angle mildly relativistic cocoon with a successful off-axis jet (e.g., \citealt{Kasliwal+etal+2017, Mooley+etal+2018b}).
The scenario consists an energetic narrow core (with Lorentz factor $ \Gamma \gtrsim 100$) and a less energetic wide-angle cocoon (with Lorentz factor $\Gamma \lesssim 10$) with angular profile of energy. The line of sight is off-axis the jet core but within the cocoon. With the decelerating of the outflow, the visible area of the outflow is increasing and including more and more energetic region. 
Hence, the observed flux will gradually rise until the emission of the jet core comes into our view entirely, and then the observed flux dropping sharply.
Moreover, the angular displacement of the radio flux centroid is consistent with the observed for this scenario (\citealt{Mooley+etal+2018b}).

GRB 170817A appears to be the case of the fourth scenario. However, by studying the observed distribution of $E_X/ E_{\gamma}$ (the ratio of the isotropic equivalent early X-ray afterglow to prompt $\gamma$-ray energy), \cite{Beniamini+Nakar+2019} found that in order to be consistent with the observations, most of observed (long) GRB should be from a narrow region around the core if the jet has angular structure. Thus GRB 170817A is a unique burst, which could be from an off-core region. Given the uniqueness of the burst, it is worth to explore other possibilities besides the angular structure model. In this paper, we consider a stratified jet (with radial speed structure) model to explain the afterglow data of GRB 170817A. The stratified jet model can also interpret the early shallow decay on timescales of $10^2-10^5$ s in X-ray afterglows. Such a radial speed structure may originate from the central engine activities \citep{Rees+Meszaros+1998, Sari+Meszaros+2000, Nakamura+Shigeyama+2006, Gill+etal+2018}. Both of the lateral spreading and non-spreading jets are considered.  The paper is organized as follows. The model and methods are described in section 2, which are including the model description, the dynamic evolution of the afterglow shock, the flux calculation, the analytical light curves from a stratified jet, and the estimation of the source size and the angular displacement of flux centroid. Fitting results and analysis are given in section 3. Finally, we make a conclusion and discussion of our results in section 4. 
%% Authors can give a citation as 'Michel et al. 1992'.
%% You may also use \cite, \citep and \citet for citation, and use Table~1 or Figure~1
%% and so forth. Using \ref and \label for cross-references of Tables/Figures
%% is a good way in adjusting/adding/removing text, tables or figures.

\section{Model and Methods}
\label{sect:Obs}

\subsection{Model}

Consider a stratified jet (hereafter 'SJ') with a radial energy distribution over the ejecta's proper velocity. The kinetic energy distribution of the ejecta is given by $E(>u)=E_0 u^{-k}$ (e.g., \citealt{Gill+etal+2018, Mooley+etal+2018a, Huang+Li+2018}, also see similar model of \citealt{Li+etal+2018}), where $u=\Gamma \beta$ is the ejecta's proper velocity, $\Gamma$ is the bulk Lorentz factor (hereafter 'LF') of the ejecta, $\beta=\sqrt{1-1/\Gamma^2}$ is the dimensionless velocity, $E_0$ is a constant, and $k$ is the energy injection index. The LF is confined in the range of $\Gamma_{\rm{min}}< \Gamma<\Gamma_{\rm{max}}$,  where $\Gamma_{\rm{min}}$ and $\Gamma_{\rm{max}}$ are the minimum and maximum LF of the stratified ejecta. The initial maximum LF is taken as a typical value of $\Gamma_{\rm{max}}=300$ and the minimum LF $\Gamma_{\rm{min}}$ is taken as a free parameter in our fitting. The reason for the maximum LF is not taken as a free parameter is that the slowest ejecta dominates the  afterglow flux and the fastest ejecta has little effect on the afterglow flux. There are two kinds of jet evolutions are investigated in the SJ model, including a lateral-spreading jet and a non-lateral-spreading jet. Note that a significant difference between the SJ and QSSEI model is that the outflow has a small initial opening angle in the SJ model, while the QSSEI model has a quasi-spherical structure.

A possible origin of the SJ is presented as follows. Multiple relativistic shells with a random distribution of LFs are ejected from the central engine, and there will be numerous collisions between the shells with different velocities. The internal shocks induced by the collisions produce the GRB prompt emission. After the numerous collisions, the distribution of shells may be formed as ordered with increasing values of LFs (e.g., \citealt{Kobayashi+etal+1997}). Subsequently, in the afterglow phase, the fastest shell gradually be decelerated by the external medium, the slower and more energetic shells will catch up with it and energize it (e.g., \citealt{Rees+Meszaros+1998, Sari+Meszaros+2000, Nakamura+Shigeyama+2006, Gill+etal+2018}).

\subsection{Dynamic evolution of the afterglow shock}

The dynamic evolution of the jet directly determine the afterglow light curve. In the ultrarelativistic phase, the jet opening angle nearly remains a constant. But the lateral expansion becomes significant when the jet is decelerated down to $\Gamma\sim 1/\theta_j$ (\citealt{Rhoads+1999}), or later (e.g.,
\citealt{vanEerten+etal+2010}), where $\theta_j$ is the half opening angle of the jet. The jet opening angle evolution can be given by (\citealt{Granot+Piran+2012})
\begin{equation}
\frac{d\theta_{j}}{dR}=\frac{1}{\Gamma^{2}\theta_{j}R}.
\end{equation}
where  $R$ is the radius of the jet. For a non-spreading jet, we have $\frac{d\theta_{j}}{dR}=0.$
The evolution of the afterglow shock's swept-up mass $m$ can be described by 
\begin{equation}\label{dmdR}
\frac{dm}{dR}=2\pi(1-\cos \theta_{j})R^2 n m_p.
\end{equation}
where $n$ is the circumburst medium number density and $m_p$ is the proton mass. We adopt a thin shell approximation in this work. In the radially structured ejecta model, $\sim 90 \%$ of the injecting energy is shared by the forward-shocked medium and $\sim 10 \%$ is shared by the reverse-shocked material, and the reverse shock is Newtonian or trans-relativistic for the thin shell approximation case (\citealt{Yu+etal+2007}). Considering the energy fraction shared by the reverse-shocked material is very small, we take an approximation that the whole injecting energy is converted into the forward-shocked medium. Thus we obtain (\citealt{Huang+Li+2018})
\begin{equation}\label{mgamma} 
mc^2 [g(\Gamma)-1] \approx E(>\Gamma \beta)=E_0(\Gamma \beta)^{-k}.
\end{equation}
where $g(\Gamma)=\frac{\Gamma^3+(\hat{\gamma}-1)(\Gamma^2-1)(\Gamma-1)}{\Gamma}$, $\hat{\gamma}=(4\Gamma+1)/(3\Gamma)$ is the adiabatic index of the shocked medium, $c$ is the light speed, and $E(>\Gamma \beta)=E_0(\Gamma \beta)^{-k}$ is the kinetic energy of ejecta. Note that equation 3 is available for both relativistic and Newtonian cases. Combining equations 1, 2 and 3, we can obtain the evolution of $\theta_j$, $m$ and $\Gamma$ over $R$.

\subsection{Flux calculation}

The observed flux for a THJ at a given observed time $T$ can be given by (\citealt{Granot+etal+1999, Woods+Loeb+1999})
\begin{equation}
F_{\nu}(T)=\frac{1+z}{D_{L}^{2}}\int_{0}^{\theta_{v}+\theta_{j}} 2\Delta \phi (\theta_{v},\theta_{j},\theta)R^{2} \delta^{3} I'_{\nu'} \cos \theta \sin \theta d\theta.
\end{equation}
where $D_L$ is the luminosity distance of the burst, $z$ is the redshift, $\delta=1/\Gamma(1-\beta \cos \theta)$ is the Doppler factor, $\theta_v$ is the viewing angle, and $\theta$ is the angle between the speed of a volume element within the jet and the viewing line. 
The observed time $T$ is a function of $\Gamma$, $R$, and $\theta$, which is
\begin{equation}\label{obsT}
    \frac{T}{1+z}=\int_0^R\frac{dr}{\beta c}-\frac{R\rm{cos}\theta}{c}.
\end{equation}
$\Delta \phi$ is given by
\begin{equation}
\Delta \phi= 
\begin{cases}
0 \qquad \qquad \qquad \qquad  \qquad  &(\theta_{v}>\theta_{j}, \theta<\theta_{v}-\theta_{j}) \\
\pi  & (\theta_{v}<\theta_{j}, \theta<\theta_{j}-\theta_{v}) \\
\cos^{-1}(\frac{\cos \theta_{j} -\cos \theta \cos \theta_{v}}{\sin \theta_{v} \sin \theta}) &(\rm{otherwise}).
\end{cases}
\end{equation}
$I'_{\nu'} =j'_{\nu'}\Delta' $ is the radiation intensity in the jet comoving frame, and $j'_{\nu'}=\frac{N_{e}}{4 \pi R^{2}\Delta'} \frac{P'_{\nu',max}}{4 \pi}f(\nu')$  is the comoving emissivity. 
$\Delta'$ is the comoving width of the shocked material and $N_{e}=4 \pi R^{3} n /3$ is the total number of swept-up electrons in the shocked fluid.
$P'_{\nu',max}=\phi_p \frac{\sqrt{3} q_{e}^3 B}{m_e c^2}$ (\citealt{Wijers+Galama+1999}) and $f(\nu')$ are respectively the comoving peak spectral power and dimensionless spectrum in the comoving frame, where $\phi_p$ is the dimensionless peak flux and $q_e$ is the electron charge. $B=\sqrt{8\pi n m_p c^2 \epsilon_B (\Gamma-1)(4\Gamma+3)}$ (\citealt{Sari+Piran+1995, Sari+etal+1998}) is the magnetic field strength of the comoving frame, where $\epsilon_B$ is the energy fraction of the magnetic field. For the fast cooling and slow cooling cases, $f(\nu')$ are respectively given by (\citealt{Sari+etal+1998})
\begin{equation}
f(\nu')=
\begin{cases}
(\nu'/\nu'_{c})^{1/3}  \qquad \qquad \qquad &(\nu'<\nu'_c)  \\
(\nu'/\nu'_c)^{-1/2} &(\nu'_c<\nu'<\nu'_m) \\
(\nu'_m/\nu'_c)^{-1/2}(\nu'/\nu'_m)^{-p/2} &(\nu'>\nu'_m),
\end{cases}
\end{equation}
and 
\begin{equation}
f(\nu')= 
\begin{cases}
(\nu'/\nu'_{m})^{1/3}  \qquad \qquad \qquad &(\nu'< \nu'_{m}) \\
(\nu'/\nu'_{m})^{-(p-1)/2} &(\nu'_{m}<\nu'<\nu'_{c}) \\
(\nu'_{c}/\nu'_{m})^{-(p-1)/2}(\nu'/\nu'_{c})^{-p/2}  &(\nu' > \nu'_{c}). 
\end{cases}
\end{equation}

where $\nu'_{m}=\frac{3x_p \gamma_{m}^{2} q_eB}{4\pi m_e c}$ and $\nu'_c=0.286\frac{3\gamma_{c}^{2}q_eB}{4\pi m_e c}$ are respectively the comoving synchrotron typical frequency and the cooling frequency (\citealt{Wijers+Galama+1999}). $x_p$ is the dimensionless peak frequency of the spectrum and $x_p \approx \phi_p \approx 0.6$ are adopted (\citealt{Wijers+Galama+1999}). $p$ is the electron spectrum index. $\gamma_m =\frac{m_p}{m_e}\frac{p-2}{p-1}\epsilon_e (\Gamma-1)$ and $\gamma_c \simeq \frac{6 \pi m_{e} c}{\sigma_{T} B^{2} \Gamma T}$ are respectively the minimum LF of the injected electrons and the cooling LF of the electrons, where $\epsilon_e$ is the energy fraction of the electrons, $m_e$ is the electron mass, and $\sigma_T$ is Thomson scattering cross section. Note that the synchrotron self-absorption (SSA) is not considered in our calculation, since the rising phase of the observed radio, optical, and X-ray light curves of GRB 170817A are in the same spectral segment. The observed flux thus can be written as 
\begin{equation}
F_{\nu}(T)=\frac{1+z}{6\pi D_{L}^{2}}\int_{max (\theta_{v}-\theta_{j},0)}^{\theta_{v}+\theta_{j}} \Delta \phi (\theta_{v},\theta_{j},\theta)R^{3}(T,\Gamma)
n \delta^{3} f(\nu') P'_{\nu',max} \cos \theta \sin \theta d\theta. 
\end{equation}

\subsection{The analytical light curve scalings from a stratified jet}

 We attempt to give the analytical light curve from a SJ with the radial profile of energy. The circumburst medium (CBM) number density $n$ is considered as a constant in this paper. We first consider a spherical outflow. The swept up mass of the external shock is $m=\frac{4}{3}\pi R^3 n m_p$.
For the ultrarelativistic case, we have $\Gamma \gg 1$ and $\beta \sim 1$. Thus we get $g(\Gamma)-1= u^2(1+\frac{\beta^2}{3}) \approx \frac{4 u^2}{3} \approx \frac{4\Gamma^2}{3}$, and combining equation 3, we obtain
\begin{equation} \label{gama-t}
\Gamma=[\frac{9E_0}{128 \pi T^3 n m_p c^5}]^{\frac{1}{(8+k)}} \propto T^{\frac{-3}{(8+k)}},
\end{equation}
where the observed time $T \approx R/2 \Gamma^2 c$ is used.
We can find the scalings: $R \propto T^{\frac{2+k}{8+k}}$, $B \simeq \Gamma \sqrt{32\pi n m_p c^2 \epsilon_B} \propto \Gamma$, $\gamma_m \propto \Gamma$, $\gamma_c \propto \Gamma^{-3} T^{-1}$, $N_e \propto R^3$, and $P_{\nu, max}\simeq \Gamma P'_{\nu',max} \propto \Gamma B \propto \Gamma^2$, where $P_{\nu, max}$ is the observed peak spectral power. Thus we obtain (\citealt{Sari+etal+1998},\citealt{Wijers+Galama+1999})
 \begin{equation} \label{numc-Fmax}
\begin{cases}
& \nu_m \simeq \Gamma \nu'_{m} \propto \Gamma \gamma_{m}^{2} B \propto \Gamma^4 \propto T^{\frac{-12}{(8+k)}} \\
& \nu_c \simeq \Gamma \nu'_{c} \propto \Gamma \gamma_{c}^{2} B \propto \Gamma^{-4} T^{-2} \propto T^{\frac{-2(2+k)}{(8+k)}} \\
& F_{\nu, max}\simeq  N_e P_{\nu, max} /4\pi D_L^2 \propto N_e P_{\nu, max} \propto R^3 \Gamma^2 \propto T^{\frac{3k}{(8+k)}}.
\end{cases}
\end{equation}
where $\nu_m$, $\nu_c$, and $F_{\nu, max}$ are respectively the synchrotron typical frequency, the cooling frequency, the peak flux in the observer frame. There are two types of spectra, depending on the order between $\nu_m$ and $\nu_c$ (\citealt{Sari+etal+1998}). For the fast cooling ($\nu_m>\nu_c$) regime, the observed flux is given by 
\begin{equation} \label{fast_col}
F_\nu\equiv F_{\nu, max}f(\nu)=F_{\nu, max}
\begin{cases}
(\nu/\nu_{c})^{1/3}   \propto T^{\frac{(11k+4)}{3(8+k)}} &(\nu<\nu_c)  \\
(\nu/\nu_c)^{-1/2}  \propto T^{\frac{2(k-1)}{(8+k)}} &(\nu_c<\nu<\nu_m) \\
(\nu_m/\nu_c)^{-1/2}(\nu/\nu_m)^{-p/2} \propto T^{\frac{-(k+6p-4)}{(8+k)}} &(\nu>\nu_m),
\end{cases}
\end{equation}
while for the slow cooling ($\nu_c>\nu_m$) regime, the observed flux is 
\begin{equation} \label{slow_col}
F_{\nu}= 
F_{\nu, max}\begin{cases}
(\nu/\nu_{m})^{1/3} \propto T^{\frac{(4+3k)}{(8+k)}} &(\nu< \nu_{m}) \\
(\nu/\nu_{m})^{-(p-1)/2}  \propto T^{\frac{3(2-2p+k)}{(8+k)}}  &(\nu_{m}<\nu<\nu_{c}) \\
(\nu_{c}/\nu_{m})^{-(p-1)/2}(\nu/\nu_{c})^{-p/2}  \propto T^{\frac{-(k+6p-4)}{(8+k)}}  &(\nu > \nu_{c}).
\end{cases}
\end{equation}
where $f(\nu)$ is the dimensionless spectrum in the observer frame. The above analytical results are based on the assumption of a spherical outflow, but we consider a collimated jet in this paper. Actually, if the following two conditions are met, the above analytical results are still applicable to the jet case. The first is the lateral expansion is not important, and the second is the jet edge is not seen ($1/\Gamma<\theta_v+\theta_j$). If the lateral expansion is not important but the jet edge is seen ($1/\Gamma>\theta_v+\theta_j$), then the flux in eq. \ref{numc-Fmax}, \ref{fast_col} and \ref{slow_col}  should be multiplied by a reduction factor of $\theta_j^2/(1/\Gamma)^2=\Gamma^2 \theta_j^2$ (e.g., \citealt{Panaitescu+etal+1998}; \citealt{Gao+etal+2013}). Thus we get $F_{\nu, max} \propto R^3 \Gamma^4 \propto T^{\frac{(3k-6)}{(8+k)}}$, and the observed flux for the fast and slow cooling regimes can be respectively given by
\begin{equation} \label{fast_col2}
F_\nu =F_{\nu, max}
\begin{cases}
(\nu/\nu_{c})^{1/3}   \propto T^{\frac{(11k-14)}{3(8+k)}} &(\nu<\nu_c)  \\
(\nu/\nu_c)^{-1/2}  \propto T^{\frac{2(k-4)}{(8+k)}} &(\nu_c<\nu<\nu_m) \\
(\nu_m/\nu_c)^{-1/2}(\nu/\nu_m)^{-p/2} \propto T^{\frac{-(k+6p+2)}{(8+k)}} &(\nu>\nu_m),
\end{cases}
\end{equation}
and
\begin{equation} \label{slow_col2}
F_{\nu}= 
F_{\nu, max}\begin{cases}
(\nu/\nu_{m})^{1/3} \propto T^{\frac{(3k-2)}{(8+k)}} &(\nu< \nu_{m}) \\
(\nu/\nu_{m})^{-(p-1)/2}  \propto T^{\frac{3(k-2p)}{(8+k)}}  &(\nu_{m}<\nu<\nu_{c}) \\
(\nu_{c}/\nu_{m})^{-(p-1)/2}(\nu/\nu_{c})^{-p/2}  \propto T^{\frac{-(k+6p+2)}{(8+k)}}  &(\nu > \nu_{c}).
\end{cases}
\end{equation}

Note that the pure edge effect has no effect on the Newtonian phase since $\Gamma \sim 1$ in this phase. Given $\Gamma \sim 1$ and $\beta \ll 1$, we can get $g(\Gamma)-1=u^2(1+\frac{\beta^2}{3}) \approx u^2 \approx \beta^2$, and combining equation 3, we obtain
%
%\begin{equation}\label{mgamma} 
%\add{mc^2 [g(\Gamma)-1] \approx mc^2 \beta^2 \approx E_0 \beta^{-k}.}
%\end{equation}
%
\begin{equation}
\beta \approx [\frac{3E_0}{4\pi R^3 n m_p c^2}]^{\frac{1}{(2+k)}} \propto R^{\frac{-3}{(2+k)}}.
\end{equation}
 Considering $R \approx \beta c T$, one can find the scalings: $\beta \propto T^{\frac{-3}{(5+k)}}$, $R \propto T^{\frac{2+k}{5+k}}$, $B \propto \beta \propto T^{\frac{-3}{(5+k)}}$, $\gamma_m \propto \beta^2 \propto T^{\frac{-6}{(5+k)}}$, $\gamma_c \propto B^{-2}T^{-1} \propto T^{\frac{(1-k)}{(5+k)}}$, $\nu_m \propto \gamma_{m}^{2}B \propto T^{\frac{-15}{(5+k)}}$, $\nu_c \propto \gamma_{c}^{2}B \propto T^{\frac{-(1+2k)}{(5+k)}}$, and $F_{\nu, max} \propto N_e P_{\nu, max} \propto R^3 B \propto T^{\frac{3(1+k)}{(5+k)}}$. Thus the observed flux in the Newtonian phase can be given by
\begin{equation} \label{newtonian_ph}
F_{\nu}= 
\begin{cases}
(\nu/\nu_{m})^{-(p-1)/2} F_{\nu, max} \propto  T^{\frac{3(2k-5p+7)}{2(5+k)}}  &(\nu_{m}<\nu<\nu_{c}) \\
(\nu_{c}/\nu_{m})^{-(p-1)/2}(\nu/\nu_{c})^{-p/2} F_{\nu, max}  \propto T^{\frac{(4k-15p+20)}{2(5+k)}}  &(\nu > \nu_{c}).
\end{cases}
\end{equation}
These results, if there is no energy injection or energy injection have ceased, are consistent with those derived by some authors \citep{Dai1999ApJ...519L.155D,Frail2000ApJ...537..191F,Livio2000ApJ...538..187L,Huang2003MNRAS.341..263H,zhang_dynamics_2009}.

For more later times, $\gamma_m<2$ is reached, below which the synchrotron approximation becomes invalid. We can neglect the emission from electrons with $\gamma_e<2$. Using the treatment of \cite {2006ApJ...638..391G}, the total electron number emitting synchrotron photons $N_e\propto \beta^2R^{3}$ and $\gamma_m=2$ (or see the similar treatment in \citealp{2013ApJ...778..107S} and \citealp{2018ApJ...859..123H}), we can find 
\begin{equation} \label{gm<2_newtonian}
F_{\nu}=
\begin{cases}
(\nu/\nu_m)^{-(p-1)/2}F_{\nu,max} \propto T^{\frac{3(2k-p-1)}{2(5+k)}}  &\nu_m<\nu<\nu_c,\\
(\nu/\nu_c)^{-p/2}(\nu_c/\nu_m)^{-(p-1)/2}F_{\nu,max}  \propto T^{\frac{(4k-3p-4)}{2(5+k)}}  \qquad \qquad &\nu>\nu_c. 
\end{cases}
\end{equation}

When $\epsilon_e$ is small, it is possible that $\gamma_m<2$ arrives earlier than the Newtonian phase of the bulk LF (see the fitting results of GRB 170817A in the next section). In this situation, the velocity of the jet is relativistic, while $\gamma_m<2$ is reached. If the jet edge is seen, the peak flux should be corrected as $F_{\nu,max} \propto \Gamma^2 (\Gamma-1)N_e P_{\nu, max}\approx \Gamma^3N_e P_{\nu, max} \propto T^{\frac{3(k-1)}{(8+k)}}$, where the factor $\Gamma-1$ is taking into account the fraction of the relativistic electrons ($\gamma_e>2$) in the total electrons. Thus the observed flux can be given by
\begin{equation} \label{gm<2_rel}
F_{\nu}=
\begin{cases}
(\nu/\nu_m)^{-(p-1)/2}F_{\nu,max}, \propto T^{\frac{3(-2+k-p)}{(8+k)}}  &\nu_m<\nu<\nu_c,\\
(\nu/\nu_c)^{-p/2}(\nu_c/\nu_m)^{-(p-1)/2}F_{\nu,max} \propto T^{\frac{(2k-3p-2)}{(8+k)}}  \qquad \qquad &\nu>\nu_c. 
\end{cases}
\end{equation}
Note that all the scalings in this section are applicable to the case of on-axis ($\theta_v \lesssim \theta_j$) observation and the case of $\theta_v - \theta_j \lesssim \frac{1}{\Gamma}$.
\subsection{Estimation of the source size and angular displacement of flux centroid}
 We estimate the observed angular size and the angular displacement of flux centroid of GRB afterglow image by establishing the following coordinate system (e.g., \citealt{Gill+Granot+2018b}). The jet symmetry axis is selected as the $z$-axis, while the line of sight is the $\tilde{z}$-axis and is in the $x-z$ plane. The $y$ and $\tilde{y}$ axes coincide. The observed image lies in the $\tilde{x}- \tilde{y}$ plane. The observed image has two mutually perpendicular scales, where $R_{\perp \tilde{x}}$ is the size perpendicular to the line of sight and is parallel to $\tilde{x}$-axis, while $R_{\perp \tilde{y}}$ is the size perpendicular to the line of sight and is parallel to $\tilde{y}$-axis. For a given observed time $T$, $R_{\perp \tilde{x}}$  can be estimated as
\begin{equation}
R_{\perp \tilde{x}} \approx
\begin{cases}
R_g \sin{(\frac{1}{\Gamma})} & (\theta_j - \theta_v \gtrsim \frac{1}{\Gamma})\\
[R_g \sin{(\frac{1}{\Gamma})} - R_{e} \sin{(\theta_v -\theta_j)}]/2 & (0< \theta_v - \theta_j \lesssim \frac{1}{\Gamma}),\\
\end{cases}
\end{equation}
where $R_g$ and $R_e$ are the shock radii on $\theta=1/ \Gamma$ and on the jet edge of $\theta=\theta_v - \theta_j$ at the observed time $T$, respectively. 
The radii can be derived by using the dynamic evolution of $\Gamma-R$ relation and eq. \ref{obsT}. $R_{\perp \tilde{y}}$ can be estimated by
\begin{equation}
R_{\perp \tilde{y}}\approx 
\begin{cases}
R_g \sin{(\frac{1}{\Gamma})} & (\theta_j - \theta_v \gtrsim \frac{1}{\Gamma})\\
R_{g} \sin{(\frac{1}{\Gamma})} \sin{\Delta \phi_g} & (\theta_v > \theta_j, \theta_v - \theta_j <\frac{1}{\Gamma} < \theta_m) \\
R_{m} \sin{\theta_m} \sin{\Delta \phi_m}  & (\theta_v >\theta_j, \theta_m \leq \frac{1}{\Gamma}),\\
\end{cases}
\end{equation}
where $\Delta \phi_g = \cos^{-1}(\frac{\cos{\theta_{j}} -\cos{(\frac{1}{\Gamma})}  \cos{\theta_{v}}}{\sin{\theta_{v}} \sin{(\frac{1}{\Gamma})}})$, $\Delta \phi_m = \cos^{-1}(\frac{\cos{\theta_{j}} -\cos{\theta_m} \cos{\theta_{v}}}{\sin{\theta_{v}} \sin{\theta_m}})$, $\theta_m=\cos^{-1}(\cos{\theta_j} \cos{\theta_v})$, and $R_m$ is the shock radius on $\theta=\theta_m$ at the observed time $T$. Hence, the observed angular sizes at the two mutually orthogonal directions are given by
\begin{equation}
\begin{cases}
\theta_{\perp \tilde{x}} \approx \frac{R_{\perp \tilde{x}}}{D_A} \approx (1+z)^2 R_{\perp \tilde{x}}/D_L \\
\theta_{\perp \tilde{y}} \approx \frac{R_{\perp \tilde{y}}}{D_A} \approx (1+z)^2 R_{\perp \tilde{y}}/D_L.
\end{cases}
\end{equation}
where $D_A =D_L/(1+z)^2$ is the angular distance of the burst. The flux centroid is moving along the $\tilde{x}$-axis, and by taking an approximation, we obtain
\begin{equation} \label{flux_mo}
\tilde{x}_{fc}(T) \approx 
\begin{cases}
0  & (\theta_j - \theta_v \gtrsim \frac{1}{\Gamma})\\
R_v \sin{\theta_v} & (\theta_v > \theta_j,\theta_v + \theta_j \lesssim \frac{1}{\Gamma})\\
R_{e} \sin{(\theta_v -\theta_j)}  & (\theta_v > \theta_j, \theta_v -\theta_j > \frac{1}{\Gamma}) \\
[R_g \sin{(\frac{1}{\Gamma})} + R_{e} \sin{(\theta_v -\theta_j)}]/2  & (\theta_v > \theta_j, \theta_v - \theta_j < \frac{1}{\Gamma},\theta_v + \theta_j \gg \frac{1}{\Gamma})\\
\end{cases}
\end{equation}
where $R_v$ is the shock radius on the jet symmetry axis of $\theta=\theta_v$ at the observed time $T$. Thus the angular displacement and the apparent velocity of the flux centroid are respectively given by
\begin{equation}
\begin{cases}
\Delta \theta_{fc} \approx \frac{|\tilde{x}_{fc}(T_2) - \tilde{x}_{fc}(T_1)|}{D_A} \approx (1+z)^2 |\tilde{x}_{fc}(T_2) - \tilde{x}_{fc}(T_1)|/D_L \\
\beta_{app} \approx \frac{|\tilde{x}_{fc}(T_2) - \tilde{x}_{fc}(T_1)|}{c(T_2 - T_1)}
\end{cases}
\end{equation}
\section{Fitting results and analysis}

We use a package MceasyFit in the light curve fitting. The package is based on the Markov-Chain Monte Carlo (MCMC) method, which is described in \cite{Zhang+etal+2016}. We fit 3 GHz and 5.5 GHz radio, optical $5.1 \times 10^{14}$ Hz (F606W), and X-ray (1keV) afterglow light curves. The afterglow data in the four bands are taken from the following papers:
\cite{Hallinan+etal+2017, Alexander+etal+2018, Dobie+etal+2018, Margutti+etal+2018, Mooley+etal+2018a, Mooley+etal+2018c, Piro+etal+2019, Lyman+etal+2018, Lamb+etal+2019, Nynka+etal+2018, Troja+etal+2018, Troja+etal+2019, Troja+etal+2020}, and \cite{Hajela+etal+2019}. There are nine free parameters in our fitting, i.e.,
\begin{equation}
\Phi=[\theta_0, \theta_v, \epsilon_B, \epsilon_e, E_1, n, \Gamma_{\rm{min}}, k, p].
\end{equation}
where $\theta_0$ is the initial half opening angle of the jet, and $E_1$ is a constant, which is related to $E_0$ by $E_0=E_1 \times (1- \cos \theta_0)$. In the fitting, we confined the free parameters as $\theta_0 \in [0.01,0.45]$, $\theta_v \in [0,0.6]$, $log_{10}(\epsilon_B) \in [-7,-0.1]$, $log_{10}(E_1) \in [53,66]$, $log_{10}(n)\in [-7.5,1]$, $\Gamma_{\rm{min}} \in [2,10]$, $k \in [2,10]$, $p \in [2.05,2.6]$, and $log_{10}(\epsilon_e) \in [-6,-0.1]$. Note that the viewing angle is constrained to $\theta_v > \theta_0$ in the fitting since the superluminal motion of the centre-of-brightness on the sky in GRB 170817A was observed (\citealt{Mooley+etal+2018b}).

We use the SJ models to fit the afterglow datasets of GRB 170817A, and the best fit parameters for the spreading and the non-spreading cases are displayed in Table 1. The 3 GHz and 5.5 GHz radio, optical (F606W), and X-ray (1keV) light curves derived with the best fit parameters are shown in Fig. 1 and Fig. 2 for the spreading jet and non-spreading jet, respectively. The corner plots for the spreading jet and non-spreading jet are respectively shown in Fig. 3 and Fig. 4. The dynamic evolutions of the spreading and non-spreading jets are shown in Fig. 5.
As shown in Fig. 1 and Fig. 2, both the models can give good fits to the afterglow datasets of GRB 170817A with $\rm{\chi^2/DOF \approx 1.47}$ and $\approx 1.38$ for the non-spreading and spreading jets, respectively. As shown in Table 1, $\theta_0 \approx 8.3^{\circ}$ and $\theta_v \approx 9.9^{\circ}$ for non-spreading jet, and $\theta_0 \approx 5.7^{\circ}$ and $\theta_v \approx 8.2^{\circ}$ for spreading jet. These results suggest that both of these two scenarios are slightly off-axis observed at the beginning of the afterglow. The energy injection indexes take the values of $k\simeq 7.1$ and $k\simeq 8.4$ for non-spreading and spreading jets, respectively. The true energy of the non-spreading and spreading jets are respectively given by $E_{\rm{jet}}\simeq E_{\rm{max}}=E_0 (\Gamma_{\rm{min}} \beta_{\rm{min}})^{-k}=E_1 \times (1-\cos{\theta_0})\times (\Gamma_{\rm{min}} \beta_{\rm{min}})^{-k} \simeq 4.8 \times 10^{51}$erg and $\simeq 4.6 \times 10^{52}$erg. 

Fig. 5 shows the evolution of the LF of the fastest ejecta (upper panel), evolution of the half opening angle of the jet (middle panel), and evolution of the synchrotron typical frequency $\nu_m$ and synchrotron cooling frequency $\nu_c$ (bottom panel).  As shown in Fig. 5, both of the spreading and non-spreading jets' LF evolution curves show a change in the decline slope at about 160 days and it indicates the end of the energy injection. At the same time, the observed flux reaches its peak and begins to decrease. The evolution of $\nu_m$ and $\nu_c$ indicates the radio, optical, and X-ray emission are in the same spectral regime, i.e., slow cooling in the $\nu_{m}<\nu<\nu_{c}$ regime. Thus the light curves of the four bands in our fitting have the same slope. In the rising phase of the afterglow, the slopes of the LF evolution curves for the non-spreading and spreading jets respectively are $f \simeq - \frac{3}{8+k} \simeq -0.20$ and $\simeq -0.18$, while the analytical temporal indexes of the light curves are given by $\alpha \simeq \frac{3(2-2p+k)}{8+k} \simeq 0.9$ (see eq. \ref{slow_col}) and $\simeq 1.1$ in our fittings. The analytical temporal index $\alpha \simeq 1.1$ for the spreading jet is a little larger than $\alpha \simeq 0.8$ given by \cite{Mooley+etal+2018a}. This is due to the lateral expanding effect in the spreading jet. The jet energy will be partly spreaded laterally and be decelerated faster (see Fig. 5, even in the rising phase with small lateral expanding effect). Thus the rising slope of the light curve would be more shallow than the analytical one and the true temporal index of the rising phase can be decreased as $\alpha \simeq 0.8$. For the non-spreading jet, with the decelerates of the jet, the visible area $\theta \approx \frac{1}{\Gamma}$ has exceeded the angle of $\theta_v - \theta_j $ after $\sim 35$ days. For the spreading jet, the opening angle $\theta_j$ has already exceeded the viewing angle $\theta_v$ at $\sim 1$ day, thus the scenario has become on-axis from slightly off-axis at early rising phase. Hence, the scalings in Section 2.4 are roughly applicable for both of the non-spreading and spreading jet models in our fittings.

In the decay phase, the light curves for both spreading and non-spreading jets first experience a rapid decline from $\sim 160$ days, and then gradually transit to a shallow decline after $\sim 600$ days (see Fig. 1) and $\sim 400$ days (see Fig. 2). By numerical calculation, we found that $\gamma_m<2$ occurs at around the peak times ($\sim 160$ days) for both spreading and non-spreading jets, earlier than the arrival times of the Newtonian phase $\Gamma \sim 1$ due to the small values of $\epsilon_e$ in the two cases. The origin of the rapid decline for the spreading jet is that the lateral expansion is significant (see Fig. 5) after the light curve peaks. Although the jet edge is not seen at that time, the change of dynamic evolution due to the lateral expansion will lead to an asymptotic decay with the slope of $\alpha \sim -p \sim -2.2$ in the light curves (\citealt{Sari+etal+1999}). For the non-spreading jet, the jet edge is seen ($1/\Gamma>\theta_v+\theta_j$) and the corresponding analytical temporal index is $\alpha \sim \frac{3(k-p-2)}{(8+k)} \sim \frac{-3(p+2)}{8} \sim -1.6$ (see eq. \ref{gm<2_rel}). Note that eq. \ref{gm<2_rel} is obtained by assuming an ultrarelativistic ($\Gamma \gg 1$) jet. However, the LF of the non-spreading jet is $\Gamma \sim3$ at the peak time, at which the ultrarelativistic approximation $\Gamma\gg1$ is not very appropriate, so eq. \ref{gm<2_rel} is only marginally consistent with the numerical results. The reasons for the shallow decline after $\sim 600$ days and $\sim 400$ days for spreading and non-spreading jets are that the jets begin to transit to the Newtonian phase ($\Gamma < 2$) with $\alpha \sim \frac{3(2k-p-1)}{2(5+k)} \sim \frac{-3(p+1)}{10} \sim -1.0$ (see eq. \ref{gm<2_newtonian}). 

We have shown that both of the non-spreading and spreading jets can provide good fits to the afterglow light curves of GRB 170817A with plausible parameters, but the observation limits of the source angular size and the apparent velocity of the flux centroid can further constrain the two jet models. The apparent velocity of the flux centroid has been obtained as $\beta_{app} = 4.1 \pm 0.5 $ between $75$ and $230$ days from VLBI by \cite{Mooley+etal+2018b}, and the observed angular size of radio image at $207$ days is constrained as $\theta_{\perp} <2.5$ mas at $90\%$ confidence level by \cite{Ghirlanda+etal+2019}. By taking the estimation methods in section 2.5, we get $\beta_{app} \approx 3.6$ for the non-spreading jet, while $\beta_{app} \approx 0$ due to $\theta_j - \theta_v \gtrsim \frac{1}{\Gamma}$ (see eq. \ref{flux_mo}) at $75$ and $230$ days for the spreading jet. Moreover, the observed angular sizes of the source are $\theta_{\perp \tilde{x}} \approx 1.5$ mas and $\theta_{\perp \tilde{y}} \approx 2.2$ mas for the non-spreading jet, while $\theta_{\perp \tilde{x}} \approx \theta_{\perp \tilde{y}} \approx 6.3$ mas for the spreading jet. Hence, for the non-spreading jet model, both the observed angular size of the source and the apparent velocity of flux centroid meets the observation limits, while those for the spreading jet model are beyond the observation limits.
%
%               one-column-spanning table
%________________________________________ Table 2: Use_of_the routines
\begin{table}
\begin{center}
\caption[]{Best fit parameters for the SJ model for GRB 170817A }\label{Tab:publ-works}

%%Please Capitalize the First Letter of Each Notional Word in table's caption

 \begin{tabular}{r|r|r}
  \hline%\noalign{\smallskip}
  Parameter &  Spreading jet & Non-spreading jet \\
  \hline
  k & $8.42_{-0.58}^{+0.58}$ & $7.07_{-0.44}^{+0.78}$\\
  p & $2.18_{-0.01}^{+0.00}$ & $2.20_{-0.01}^{+0.01}$ \\
  $\Gamma_{\rm{min}}$  & $5.42_{-1.99}^{+0.18}$ & $3.18_{-1.11}^{+0.91}$ \\
  $\theta_{0}(^{\circ})$ & $5.71_{-4.18}^{+2.72}$ & $8.30_{-1.49}^{+3.53}$\\
  $\theta_{v}(^{\circ})$ & $8.18_{-3.70}^{+3.24}$ & $9.92_{-3.35}^{+3.27}$ \\
 $log_{10}(\epsilon_{B})$ & $-5.62_{-3.52}^{+0.30}$ & $-5.79_{-2.55}^{+0.48}$\\
 $log_{10}(\epsilon_{e})$ & $-2.79_{-0.09}^{+0.09}$ & $-2.48_{-0.72}^{+0.14}$ \\
 $log_{10}(E_{1}(erg))$ & $61.08_{-1.11}^{+1.11}$ & $57.05_{-1.34}^{+1.34}$ \\
 $log_{10}(n(cm^{-3}))$ & $-3.65_{-0.92}^{+0.92}$ & $-2.01_{-0.55}^{+2.50}$\\
 \hline%\noalign{\smallskip}
  BIC & 104.40 & 108.58 \\
  $\chi^{2}$ & 67.86 & 72.04 \\
  $\chi^{2}/DOF$ & 1.38 & 1.47 \\
  $\beta_{app}$ & 0 & 3.6 \\
  $\theta_{\perp \tilde{x}}$ (mas) & 6.3 & 1.5 \\
  $\theta_{\perp \tilde{y}}$ (mas) & 6.3 & 2.2 \\
\hline%\noalign{\smallskip}
\end{tabular}
\end{center}
\tablecomments{0.86\textwidth}{$E_1$ is not the true energy of the jet, and the true energy of the jet is given by $E_{\rm{jet}}\simeq E_{\rm{max}}=E_0 (\Gamma_{\rm{min}} \beta_{\rm{min}})^{-k}=E_1 \times (1-\cos{\theta_0})\times (\Gamma_{\rm{min}} \beta_{\rm{min}})^{-k}$. Thus the energy of the non-spreading and spreading jets respectively are $E_{\rm{jet}}\simeq 4.8 \times 10^{51}$erg and $E_{\rm{jet}}\simeq 4.6 \times 10^{52}$erg. Note that $\rm{BIC}=\chi^2 + n_p \times ln(n_d)$, where $\rm{n_p=9}$ is the number of free parameters, $\rm{n_d=58}$ is the number of data points, and $\rm{DOF=n_d - n_p =49}$ is the degree of freedom.}
\end{table}

%%%%%%%%%%%%%%%%%%%%%%%%%%%%%%%%%%%%%%%%%%%%%%%%%%%%%%%%%%%%%%
%%     Examples for figures using graphicx for LaTeX 2e
%%               -- our recommended way for embodying graphics
%%%%%%%%%%%%%%%%%%%%%%%%%%%%%%%%%%%%%%%%%%%%%%%%%%%%%%%%%%%%%%
%
%      A figure as large as the width of the column
%-------------------------------------------------------------
 \begin{figure}
   \centering
   \includegraphics[width=\textwidth, angle=0]{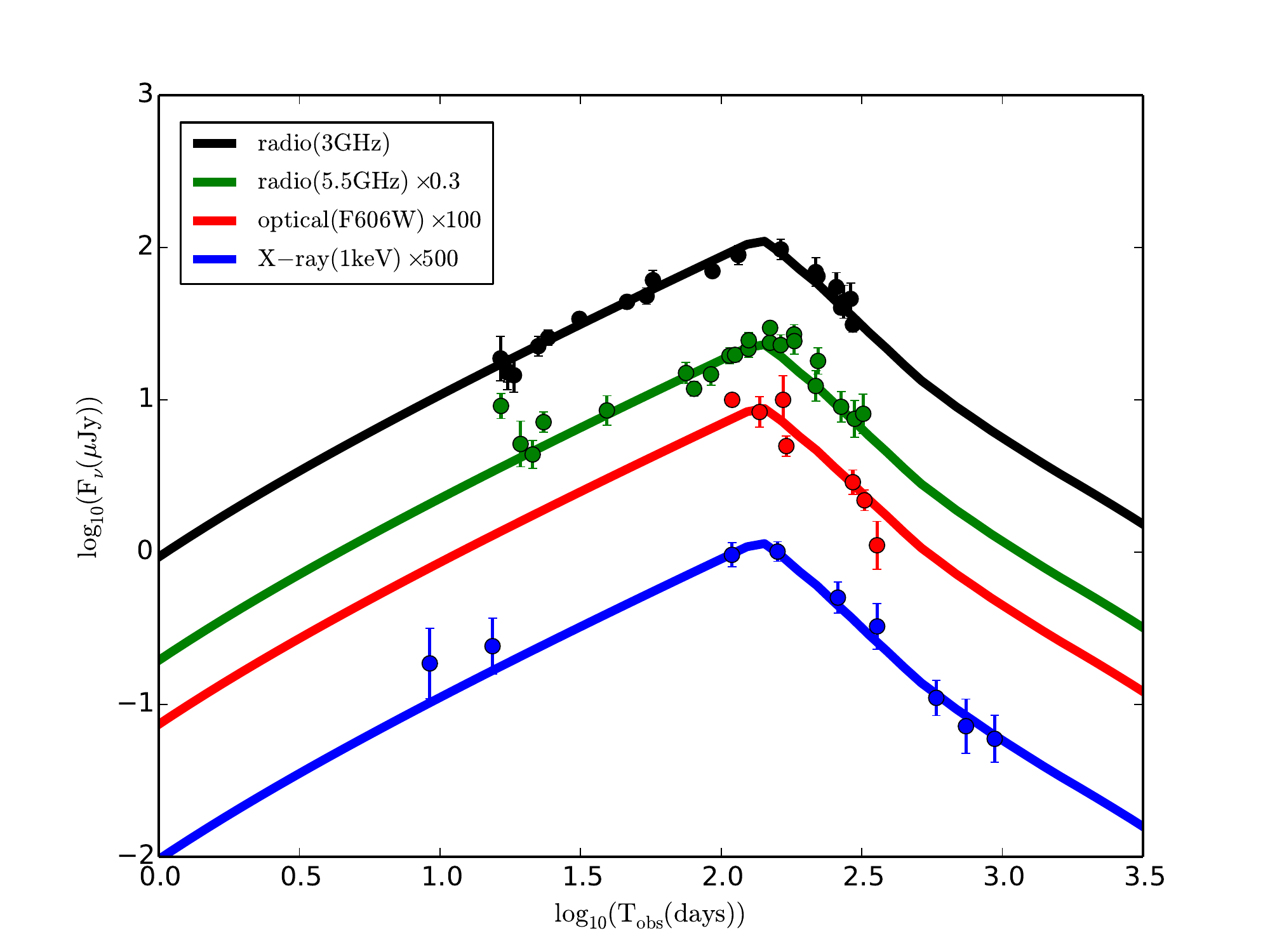}
   \caption{Multi-band afterglow light curves of the SJ model (for spreading jet) fitting to the datasets of GRB 170817A.
The solid lines show the model light curves at radio 3 GHz (black), radio 5.5 GHz (green), optical (F606W) (red), and X-ray 1 keV (blue). 
The solid circles represent the data points.}.
   \label{Fig1}
   \end{figure}
   \begin{figure}
   \centering
   \includegraphics[width=\textwidth, angle=0]{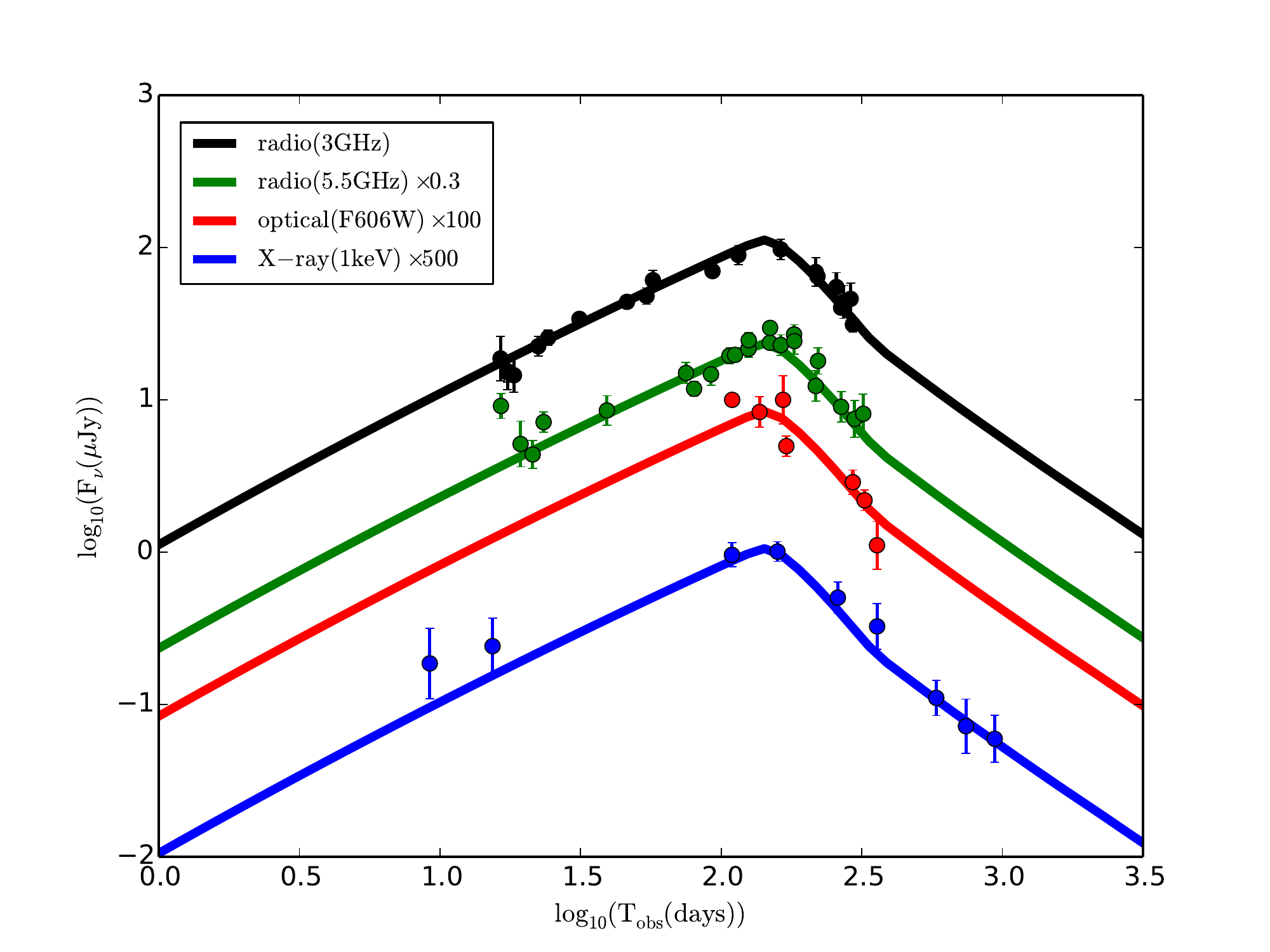}
   \caption{The same as Fig. 1, but for the non-spreading jet.}.
   \label{Fig1}
   \end{figure}
%   
%      One column rotated figure
%-------------------------------------------------------------
   \begin{figure}
   \centering
   \includegraphics[width=\textwidth, angle=0]{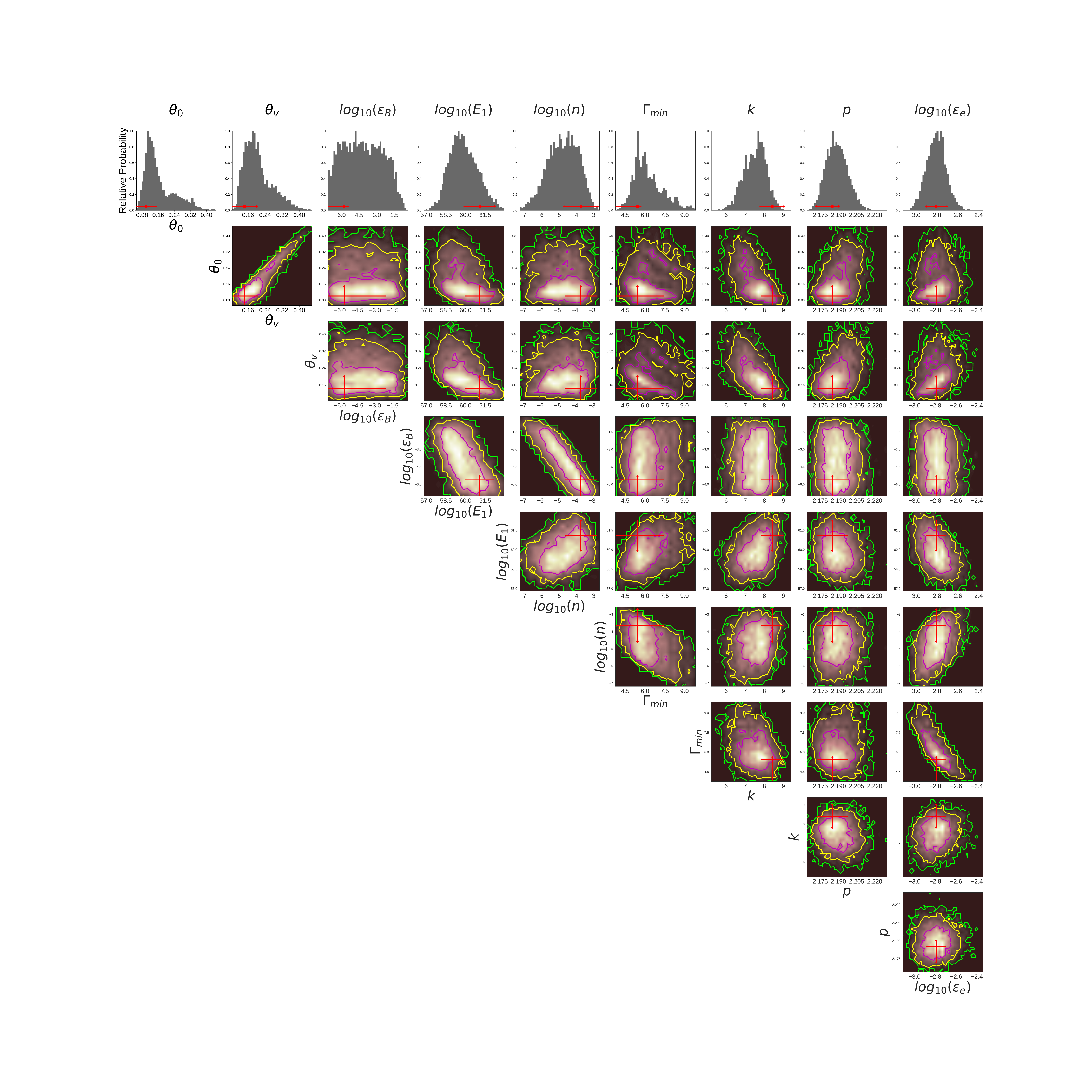}
   \caption{Corner plot showing the fitting results of the SJ model for the spreading jet.}
   \label{Fig3}
   \end{figure}
 \begin{figure}
   \centering
   \includegraphics[width=\textwidth, angle=0]{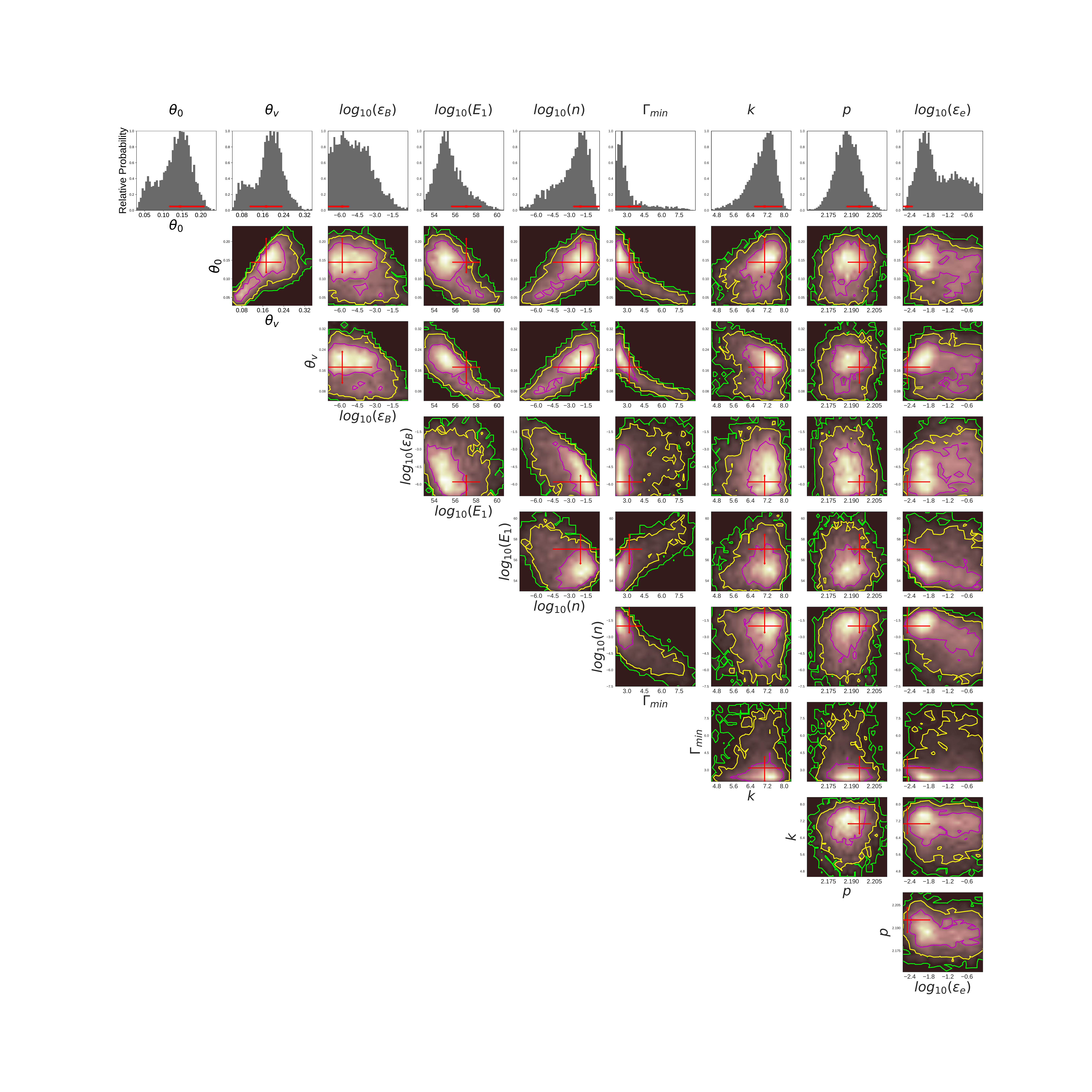}
   \caption{The same as Fig. 3, but for the non-spreading jet.}
   \label{Fig3}
   \end{figure}

%      Figure with a new BoundingBox
%-------------------------------------------------------------
\begin{figure}
\centering
\includegraphics[width=\textwidth, angle=0]{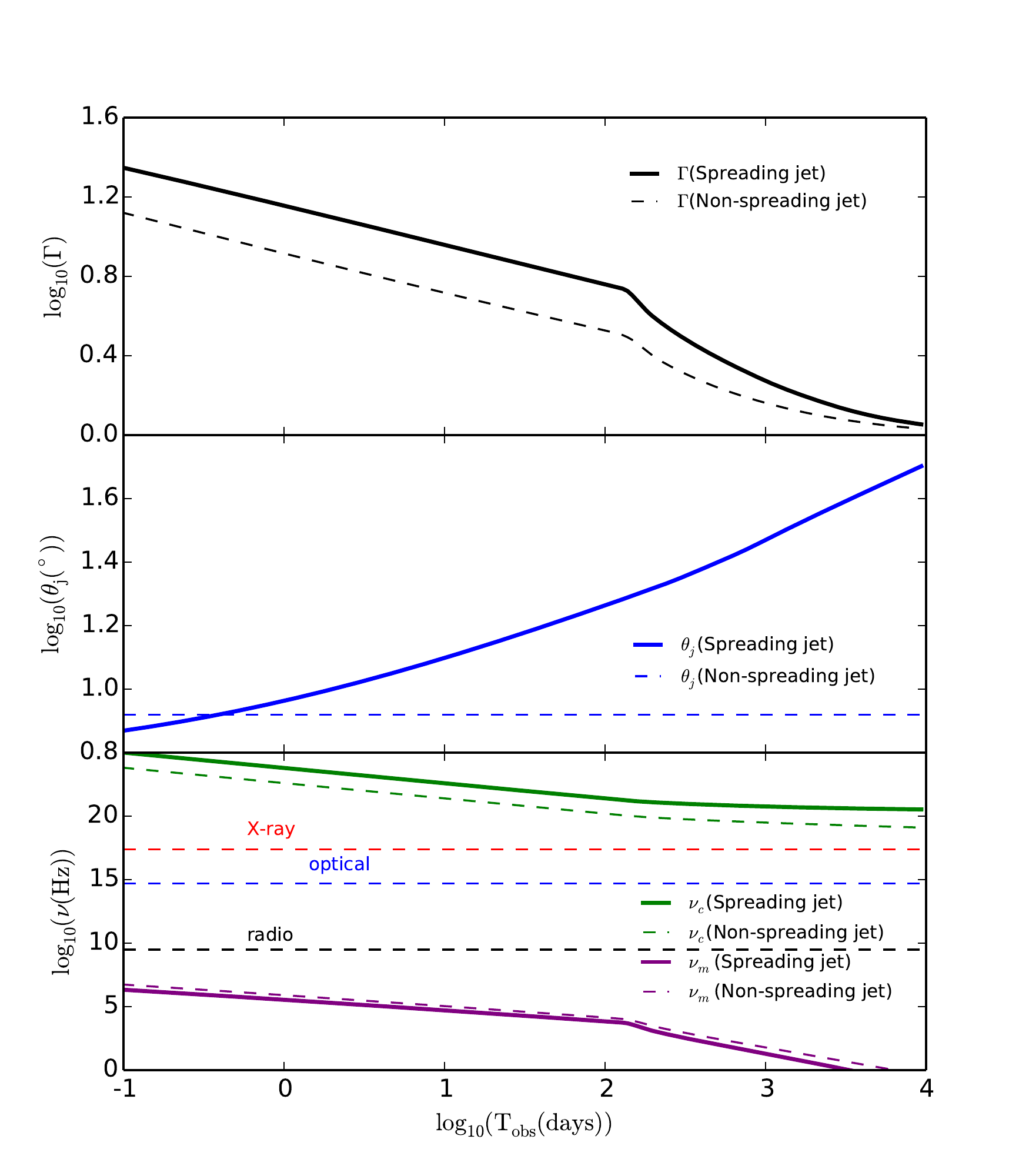}
\caption{The dynamic evolution of the SJ. The upper panel: evolution of the LF of the fastest ejecta ($\Gamma_{\rm{max}}$). The middle panel: evolution of the half opening angle of the jet. The bottom panel: evolution of the synchrotron typical frequency $\nu_m$ and synchrotron cooling frequency $\nu_c$. Note that the evolution of these parameters in the figure are from the nearest region of the jet from the viewing line, namely the region around $\theta=0$ for spreading jet and $\theta=\theta_v - \theta_0$ for non-spreading jet.}
\label{Fig:demo4}
\end{figure}

%
% one-column-wide figure(occupies half-width of a page)
%  -- This is an old way of graphics inclusion with psfig.sty
%------------------------------------------------------------ Fig1: lightcurve
%\begin{figure}
%  \vspace{2mm}
%   \begin{center}
%   \hspace{3mm}\psfig{figure=zay-f1.eps,width=80mm,height=110mm,angle=0.0}
%   \parbox{180mm}{{\vspace{2mm} }}
%   \caption{ Graphics inclusion using psfig.sty. Size controlled by the flags
%   `width' and `height'.   }
%   \label{Fig:lightcurve-ADAri}
%   \end{center}
%\end{figure}

%
%      set size-flexible figure with epsf.sty: another old way
%------------------------------------------------------------ Fig2: S/N curve
%\begin{figure}
%   \begin{center}
%   \mbox{\epsfxsize=0.6\textwidth\epsfysize=0.6\textwidth\epsfbox{zay-f2.eps}}
%   \caption{Including a figure in the manner of size-controllable using epsf.sty.
%   This figure shows a size of 0.6$\times$0.6. }
%   \end{center}
%\end{figure}

%
% two figures side-by-side with independent captions using package graphicx
% This demo can be adjusted to configure a figure with caption on left or right
%----------------------------------------------------- Figs 3 & 4:

\section{ Conclusion and discussion}
\label{sect:discussion}

In this paper, we consider a SJ model to explain the peculiar afterglow light curves of GRB 170817A and its observed image data. We investigate two kinds of jet evolutions, including a spreading jet and a non-spreading jet. By fitting the afterglow light curves of the burst in multiple bands, we find both the spreading and non-spreading jet models can give good explanation to the light curves. However, by comparing the theoretical image properties with the observations, we find the apparent angular size of the jet and the apparent velocity of the flux centroid for the non-spreading jet model can satisfy the observation limits, while those for the spreading jet model violate the observation limits. This suggests that this burst may arise from a SJ and that the spreading of the jet is not significant, at least at $\sim 230$ days post-burst. In addition, the CBM number density for the non-spreading jet model is $n\simeq 10^{-2.01}cm^{-3}$, consistent with the fact that the short burst happened in the outskirt of its host galaxy where the density should fall in between the intergalactic medium density ($\sim 10^{-6} cm^{-3}$) and the ISM number density ($\sim 1$ cm$^{-3}$). 

In the SJ model, the rising phase of the afterglow is due to the energy injection of the stratified ejecta, whose energy profile is given as $E \simeq 1.2 \times 10^{55} \times (\Gamma \beta)^{-7.1}$ for the non-spreading jet. The total kinetic energy is $E\simeq 4.8\times10^{51}$ erg and the energy injection index is $k \simeq 7.1$ for the non-spreading jet. The energy injection index is a little larger than the QSSEI model (see e.g., \citealt{Gill+etal+2018, Mooley+etal+2018a, Huang+Li+2018}). Based on our analytical results in section 2.4 (see eq. \ref{slow_col}, also see \citealt{Gill+etal+2018}), the energy injection index is given by $k \simeq (8\alpha + 6p - 6)/(3-\alpha)$ in the slow cooling phase in $\nu_{m}<\nu<\nu_{c}$ spectral regime (see eq. \ref{slow_col} and Fig. 5). The temporal index of the rising phase of GRB 170817A is $\alpha \simeq 0.8$ (\citealt{Mooley+etal+2018a}). For $\alpha \simeq 0.8$ and $p \simeq 2.2$, the energy injection index is thus $k \simeq 6.2$ (\citealt{Gill+etal+2018}). For the non-spreading jet, with the decelerates of the jet, the visible area $\theta \approx \frac{1}{\Gamma}$ has exceeded the angle of $\theta_v - \theta_j $ after $\sim 35$ days. Hence, the scalings in section 2.4 are applicable for the non-spreading jet after $\sim 35$ days. Thus the energy injection index $k \simeq 7.1$ for the non-spreading jet is close to $k \simeq 6.2$ in the QSSEI model, and the small difference comes from the jet geometry and off-axis observed scenario in the non-spreading jet model.

The steep decay of the light curves after the peak in GRB 170817A can arise from two effects. The first is the significant lateral expansion effect (e.g., \citealt{Rhoads+1999};  \citealt{Sari+etal+1999}) leads to a rapid deceleration of the jet, and thus gives rise to a steep light curve. The second is the jet edge is seen ($\frac{1}{\Gamma}>\theta_v + \theta_j $) and  the observer will feel the deficit of flux outside the jet cone compared with the spherical shock case, which leads to a steeper decay (e.g., \citealt{Panaitescu+etal+1998}). For the models in this paper, the origin of the steep decaying of the light curves in GRB 170817A after the peak for the non-spreading jet is that the jet edge is seen at $\sim 160$ days, while for the spreading jet it is that the lateral expansion is significant. The reasons for the shallow decay after $\sim 400$ days and $\sim 600$ days for non-spreading and spreading jets are that the jets begin to transit to the Newtonian phase with $\Gamma <2$. The shallow decay is consistent with the observations. It is possible that the kilonova afterglow or the energy injection by a pulsar (\citealt{Troja+etal+2020}) will contribute to the late afterglow, which will lead to a slight rise at a later time. The counter-jet can also contributes at several thousand days after the GRB trigger and will somewhat change the light curves (e.g., \citealt{Granot+etal+2018, Li+etal+2019}). Further observations in the future might reveal the origin of the shallow decay.

\begin{acknowledgements}
We thank the anonymous referee for useful suggestions and comments. This work was supported by the National Natural Science Foundation of China (Nos. U1831135,11833003), Yunnan Natural Science Foundation (2014FB188). B.B.Z acknowledges support by the National Key Research and Development Programs of China (2018YFA0404204), the National Natural Science Foundation of China (grant Nos. 11833003, U2038105), and the Program for Innovative Talents, Entrepreneur in Jiangsu. We also acknowledge the use of public data from the Fermi Science Support Center (FSSC).\end{acknowledgements}

\bibliographystyle{raa}
\bibliography{bibtex}

%\begin{thebibliography}{99}
%% you can type \apj for ApJ, \aap for A&A, \apss for Ap&SS, etc. Please consult
%% the macro chjaa.cls. You can also find them in aasguide.tex (AASTeX for ApJ, AJ, PASP)
%% Please follow the format of ChJAA's reference list

%\end{thebibliography}

%\label{lastpage}

\end{document}